\documentstyle[12pt,titlepage]{article}
\begin{document}
\pagestyle{empty}
\newcommand{\lapproxeq}{\lower
.7ex\hbox{$\;\stackrel{\textstyle <}{\sim}\;$}}
\newcommand{\skips}{\vspace{0.2cm}}
\newcommand{\skipm}{\vspace{0.4cm}}
\baselineskip=0.212in

\begin{flushleft}
\large
{MKPH-T-93-03  \hfill February 25, 1993}  \\
\end{flushleft}

\vspace{1.8cm}

\begin{center}

\Large{{\bf Tensor Structure Function b$_1$(x) }} \\

\vspace{0.3cm}

\Large{{\bf For Spin-One Hadrons}} \\

\vspace{1.8cm}

\Large
{S. Kumano $^\star$ }         \\

\vspace{1.3cm}

\Large
{Institut f\"ur Kernphysik} \\

\vspace{0.1cm}

\Large{Universit\"at Mainz}    \\

\vspace{0.1cm}

\Large
{6500 Mainz, Germany}         \\

\vspace{2.5cm}

\Large
{Proposal for the 15 GeV European Electron Facility Project} \\

\end{center}

\vspace{2.7cm}

\vfill

\noindent
{\rule{6.cm}{0.1mm}} \\

\vspace{-0.4cm}

\noindent
\normalsize
{$\star$ research supported by
         the Deutsche Forschungsgemeinschaft (SFB 201);}

\noindent
{~~E-mail: KUMANO@VKPMZP.KPH.UNI$-$MAINZ.DE.}  \\

\vfill\eject
\pagestyle{plain}
\begin{center}

\Large
{TENSOR STRUCTURE FUNCTION b$_1$(x)} \\

\Large
{FOR SPIN-ONE HADRONS} \\

\vspace{0.5cm}

{S. Kumano $^*$}             \\

{Institut f\"ur Kernphysik, Universit\"at Mainz}    \\

{6500 Mainz, Germany}      \\

\vspace{0.7cm}

\normalsize
Abstract
\end{center}
\vspace{-0.46cm}

High-energy spin physics became a popular topic
recently after the EMC finding for the proton's
spin content.
There exist unmeasured spin-dependent
structure functions ($b_1$, $b_2$, $b_3$, and $b_4$)
for spin-one hadrons such as the deuteron.
The tensor structure function $b_1(x)$
could be measured by the proposed 15 GeV
European Electron Facility. The measurement
provides important clues to physics of non-nucleonic
components in spin-one nuclei and to tensor structures
on the quark-parton level.

\vspace{0.8cm}

\noindent
{1. \underline{Introduction to $b_1$}}

\skips

Experimental results for the proton's $g_1(x)$
by the European Muon Collaboration (EMC) [1]
indicated that
``none of the proton's spin is carried by quarks''.
Since then, parton-spin distributions in the nucleon have been
a popular topic among particle and nuclear physicists.
Because of the dramatic conclusion contrary
to naive quark-model expectation, efforts
have been made to interpret it in terms of
small-$x$ contributions, gluon polarizations,
sea-quark polarizations, orbital angular momenta,
and so on [1].

Recently experimental data for $g_1(x)$ of
``the neutron'' have been taken by
the Spin Muon Collaboration (SMC) [2]
for testing the Bjorken sum rule.
Because there exists no fixed neutron target,
polarized deuterons (or $^3$He)
have to be used as the target.
However, the deuteron is interesting in its
own right.
The structure function $g_1$ exists for hadrons with
spin more than 1/2. On the other hand, the deuteron spin is one
so that other spin-dependent structure functions exist
due to its electric-quadrupole structure.
These are named $b_1$, $b_2$, $b_3$, and $b_4$
in Ref. [3].
In the Bjorken scaling limit, the only relevant structure
function is $b_1$ or equivalently $b_2$.
Detailed analyses
using the operator product expansion
and a parton model have been done
in Ref. [3], and some examples of $b_1(x)$
are also discussed.

The structure function $b_1$ can be measured by
using a target polarized parallel (and antiparallel) to
the lepton beam direction.
The lepton does not have to be polarized.
{}From the polarized cross sections and unpolarized
ones, we could obtain $b_1$.
However, because $b_1$ for a nuclear target
is considered to be very small
compared with the unpolarized one ($F_1$),
we need an intense electron-accelerator
facility to measure it.
The 15 GeV European Electron Facility provides a good
opportunity of measuring this tensor structure
function [4], which could provide clues to physics
of non-nucleonic components in
spin-one nuclei and to tensor structures
on the quark-parton level.

In section 2, we discuss structure functions for
spin-one hadrons in general. These structure
functions are expressed in terms of quark-spin
distributions by using a quark-parton model.
A phenomenological sum rule for the tensor structure
function $b_1(x)$ based on a quark-parton model
is discussed in section 3.
Examples of $b_1(x)$ are given in section 4 and
conclusions are in section 5.

$~~~$

\noindent
{2. \underline{Structure functions for spin-one hadrons}}

\skips

Structure functions for spin-one hadrons have been
investigated in detail
recently by Hoodbhoy, Jaffe, and Manohar [3].
Discussions in this section are based on their
publication. We find earlier investigations of
the tensor structure function by
Pais (for real photons) and Frankfurt-Strikman [5].

The cross section of deep-inelastic lepton scattering from
a spin-one hadron is
given by $d\sigma \propto L^{\mu\nu} W_{\mu\nu}$.
The lepton tensor is
$$
L^{\mu\nu}=k'^\mu k^\nu + k'^\nu k^\mu-g^{\mu\nu} k'\cdot k
           +i \epsilon^{\mu\nu\lambda\rho}s^e_\lambda q_\rho
{}~~~,
\eqno{(1)}
$$
where $k$ and $k'$ are incident and scattered lepton
momenta, $q$ is the momentum transfer, $s^e$ is the
electron polarization, and
$\epsilon^{\mu\nu\lambda\sigma}$ is an antisymmetric
tensor with $\epsilon_{0123}=1$.
The hadron tensor is
$$
W_{\mu \nu} (p,q,H_1,H_2)=
{1 \over {4 \pi}}
\int d^4 \xi e^{iq\cdot \xi}
<p,H_2|~[J_\mu(\xi),J_\nu(0)]~|p,H_1>
{}~~~,
\eqno{(2)}
$$
where $p$ is the hadron
momentum, $H$ and $H'$ are the $z$ components
of the hadron spin, and $J_\mu$ is the electromagnetic
current.
Using momentum conservation, parity invariance,
and time-reversal invariance, we have eight independent
amplitudes for
$\gamma (h_1)+target (H_1) \rightarrow
 \gamma (h_2)+target (H_2)$.
Therefore, the hadron tensor $W_{\mu\nu}$ can be written
in terms of eight independent structure functions:

\skipm

\noindent
$
\displaystyle{
W_{\mu \nu} = -F_1 g_{\mu \nu} +
         F_2 {{p_\mu p_\nu} \over \nu }
     + g_1 {i \over \nu}      \epsilon_{\mu \nu \lambda \sigma}
                       q^\lambda s^\sigma
{}~+~ g_2 {i \over {\nu^2}}  \epsilon_{\mu \nu \lambda \sigma}
               q^\lambda (p\cdot q s^\sigma - s \cdot q p^\sigma)
}
$

\skips

\noindent
$
\displaystyle{
{}~~~~~~~~~
{}~-~  b_1 r_{\mu \nu}    ~+~
             {1 \over 6 } b_2
            (s_{\mu \nu } + t_{\mu \nu} + u_{\mu \nu} )
{}~+~
         {1 \over 2 } b_3
            (s_{\mu \nu} - u_{\mu \nu} )
         ~+~ {1 \over 2} b_4 (s_{\mu \nu} -t_{\mu \nu})
{}~~~,
}
$
\hfill (3)

\skipm

\noindent
where $r_{\mu\nu}$, $s_{\mu\nu}$, $t_{\mu\nu}$, $u_{\mu\nu}$,
$s^\sigma$, $\nu$, and $\kappa$ are defined by

\skipm

$
\displaystyle{
{}~~~~
r_{\mu \nu} ~=~ {1 \over {\nu^2}}
              (q\cdot E^* q \cdot E  -{1 \over 3} \nu^2 \kappa )
              g_{\mu \nu}
{}~~~,
}
$
\hfill (4.1)

\skips

$
\displaystyle{
{}~~~~
s_{\mu \nu} ~=~ {2 \over {\nu^2}}
              (q\cdot E^* q \cdot E  -{1 \over 3} \nu^2 \kappa )
              {{ p_\mu p_\nu } \over \nu }
{}~~~,
}
$
\hfill (4.2)

\skips

$
\displaystyle{
{}~~~~
t_{\mu \nu} ~=~ {1 \over {2 \nu^2}}
              (q \cdot E^* p_\mu E_\nu
              +q \cdot E^* p_\nu E_\mu
              +q \cdot E   p_\mu E^*_\nu
              +q \cdot E   p_\nu E^*_\mu
              - {4 \over 3} \nu p_\mu p_\nu )
{}~,
}
$
\hfill (4.3)

\skips

$
\displaystyle{
{}~~~~
u_{\mu \nu} ~=~ {1 \over \nu}
               ( E^*_\mu E_\nu + E^*_\nu E_\mu
                + { 2 \over 3 } M^2 g_{\mu \nu}
                -{ 2 \over 3 } p_\mu p_\nu )
{}~~~,
}
$
\hfill (4.4)

\skips

$
\displaystyle{
{}~~~~
s^\sigma ~=~ - { i \over {M^2}}
        \epsilon^{\sigma \alpha \beta \tau}
           E^*_\alpha E_\beta p_\tau
{}~~~,
}
$
\hfill (4.5)

\skips

$
\displaystyle{
{}~~~~
\nu = p \cdot q ~~~,
{}~~~~~~~
\kappa =1+M^2 Q^2/\nu^2
{}~~~.
}
$
\hfill (4.6)

\skipm

\noindent
$M$ is the target hadron mass and
$E$ is the target polarization, which satisfies
$p\cdot E=0$ and $E^2 = - M^2$.
Considering current conservation, we dropped terms
proportional to $q_\mu$ and $q_\nu$ for simplicity.

Structure functions $F_1$, $F_2$,
$g_1$, and $g_2$ are defined in the same manner
as the spin-1/2 case.
The terms of new structure functions $b_{1-4}$
are symmetric
under $\mu\leftrightarrow \nu$ and under $E\leftrightarrow E^*$,
and those of $g_{1,2}$ are antisymmetric
under $\mu\leftrightarrow \nu$ and under $E\leftrightarrow E^*$.
The terms of $b_{1-4}$ vanish upon
target-spin averaging.
{}From these symmetry properties and those for $L^{\mu\nu}$
in Eq. (1), we find that
the lepton beam does not have to be polarized
for measuring $b_{1-4}$ although both polarized beam and
polarized target are necessary for $g_{1,2}$.

The new structure functions are analyzed by using
an operator product expansion [3].
The analysis shows that the twist-two contributes
to $b_1$ and $b_2$.
Because only higher orders in the twist expansion
contribute to $b_3$ and $b_4$, we do not discuss
them in this report.
There exists a
``Callan-Gross type" relation for
$b_1$ and $b_2$:
$$
b_2(x)=2x b_1(x)
{}~~~,
\eqno{(5)}
$$
which is valid only in the lowest order in QCD.
The above equation is no longer satisfied in
higher orders; however, the relations
$$
M_n(2xb_1) M_n(F_2) = M_n(b_2) M_n(2xF_1)
{}~~~~~~\rm n=odd~~~,
\eqno{(6)}
$$
are still satisfied, where the n-th moment is defined
by $\displaystyle{M_n(f)=\int x^{n-1} f(x) dx}$.
The analysis by the operator product expansion
indicates that $b_1$ and $b_2$ obey the same
scaling equations as $F_1$ and $F_2$.

We discuss twist-two structure functions,
$F_1$, $F_2$, $g_1$,
$b_1$, and $b_2$, for spin-1 hadrons
in a parton model [3].
The hadron tensor $W_{\mu\nu}$ is obtained
for a lepton scattering from free quarks.
Comparing the results with Eq. (2), we write
$F_1(x)$, $g_1(x)$, and $b_1(x)$
in terms of quark (spin) distributions
in the hadron as

\skipm

$
{}~~~~~~~~
F_1 (x) ~=~ {1 \over 2}
          {\displaystyle \sum_i} e_i^2~ [~q_i(x)+\bar q_i(x)~]
{}~~~,
$
\hfill (7.1)

\skips

$
{}~~~~~~~~
g_1 (x) ~=~
          {\displaystyle \sum_i} e_i^2~
            [~\Delta q_i(x)+ \Delta \bar q_i(x)~]
{}~~~,
$
\hfill (7.2)


$
{}~~~~~~~~~~~~~~
\Delta q_i (x) ~=~  {1 \over 2}
          [ {q_{\uparrow i} ^{+1} } (x)
          - {q_{\downarrow i} ^{+1}} (x) ]
{}~~~,
$
\hfill (7.3)

\skipm

$
{}~~~~~~~~
b_1 (x) ~=~
          {\displaystyle \sum_i} e_i^2~
            [~\delta q_i(x)+ \delta \bar q_i(x)~]
{}~~~,
$
\hfill (7.4)


$
{}~~~~~~~~~~~~~~
\delta q_i (x) ~=~  {q_\uparrow ^0} _i (x)
          - {1 \over 2}
          [ {q_{\uparrow i} ^{+1} } (x) + {q_{\uparrow i} ^{-1}} (x) ]
           ~=~  {1 \over 2} [{q_i ^0}  (x)
                            -{q_{i} ^{+1} } (x)]
{}~~~.
$
\hfill (7.5)

\skipm

\noindent
$F_2$ structure function is given by the Callan-Gross
relation $F_2(x)=2x F_1(x)$ and $b_2$ is in the similar equation
$b_2(x)=2x b_1(x)$.
As it is shown above, the $b_1(x)$ does not depend on the
quark spin but on the hadron one. It is very different
from the $g_1$ structure function, hence it probes
different spin structures.

$~~~$

\noindent
{3. \underline{Sum rule for $b_1(x)$ in a parton model}}

\skips

We discuss a sum rule for the $b_1$ structure function
in a parton model. The following discussions are
based on the derivation by Close and Kumano [6].
It should be noted that the sum rule is not a ``strict'' one
such as those derived by current algebra [7].
It is rather a phenomenological
sum rule based on a naive parton model. This is because
an assumption for sea-quark tensor polarizations
needs to be introduced in order to reach
the sum rule.
The situation is very similar to the Gottfried sum rule,
which became an interesting topic recently due
to the results obtained by the New Muon Collaboration (NMC) [8].
The $SU(2)_{flavor}$ symmetric sea
has to be assumed in order to get the
Gottfried sum rule. Therefore,
it is also not a ``strict" sum rule, but it is the one
based on a naive parton model.
Nevertheless, as the Gottfried sum rule became an important
topic for investigating the $SU(2)_f$ breaking
in antiquarks, the $b_1$ sum rule could become useful
for studying tensor polarizations in sea quarks.

Integrating Eq. (7.4) for the deuteron over $x$, we have
$$
I(b_1^D)\equiv {\displaystyle \int} dx  b_1^D (x)
          = {\displaystyle \int} dx
          [ {4 \over 9} (\delta u^D+\delta \bar u^D)
           +{1 \over 9} (\delta d^D+\delta \bar d^D)
           +{1 \over 9} (\delta s^D+\delta \bar s^D) ]
{}~~~.
\eqno{(8)}
$$
The valence distribution in the deuteron is defined by
$q_v^D= q^D- \bar q^D$, which
obviously comes from the valence quarks
in the proton and neutron:
$u_v^D= u_v^p+ u_v^n= u_v+ d_v$,
$d_v^D= d_v^p+ d_v^n= d_v+ u_v$.
Then, the above equation becomes
$$
{}~~~~~~~ I(b_1^D) = {5 \over 9} {\displaystyle \int} dx
                 [\delta u_v(x)+\delta d_v(x)]
      +{1 \over {9}} \delta Q_{sea}^D
{}~~~,
\eqno{(9)}
$$
where
$$
\delta Q_{sea}^D= {\displaystyle \int} dx
          [8 \delta \bar u (x) +2 \delta \bar d (x)
           +\delta s (x) +\delta \bar s (x)]^D
{}~~~.
\eqno{(10)}
$$
In a naive parton model, there is no tensor polarization
in sea quarks: $\delta Q_{sea}$=0.

As discussed in Ref. [9], we try to relate
the right hand sides of Eq. (9) to
the following elastic amplitude
$$
\Gamma _{H,H} ~=~ <p,H ~|~ J_0 (0) ~|~ p,H>
{}~~~.
\eqno{(11)}
$$
We calculate the above amplitude in the infinite
momentum frame in order to use a quark-parton picture.
The amplitudes is then described
in terms of quark distributions in the hadron as
$$
\Gamma_{H,H} = {\displaystyle \sum_i e_i \int } dx
                        [ q_{\uparrow i}^{H} (x)
                         +q_{\downarrow i}^{H} (x)
                         -\bar q_{\uparrow i}^{H} (x)
                         -\bar q_{\downarrow i}^{H} (x)]
{}~~~.
\eqno{(12)}
$$
The tensor combination of the amplitudes is
expressed by
$
\delta q_i^D (x) -\delta \bar q_i^D (x)
$.
Because $q_i^D - \bar q_i^D$ is the
valence quark in the deuteron and
$q_v^D= q_v^p+ q_v^n$, we obtain
$$
{1 \over 2 }~ [ \Gamma _{00} -
                       {1 \over 2}( \Gamma _{11} +\Gamma_{-1-1})]
          = {1 \over 3} {\displaystyle \int} dx
          ~[~\delta u_v (x)+\delta d_v (x)~]
{}~~~.
\eqno{(13)}
$$
The right hand side is identical to the first term
in Eq. (9), so that the integral of
$b_1$ is written by the elastic
amplitudes as
$$
I(b_1^D)
         ={5 \over 6} [ \Gamma_{00} - {1 \over 2}
                       (\Gamma_{11}+\Gamma_{-1-1})]
           +{1 \over 9} \delta Q_{sea}
{}~~~.
\eqno{(14)}
$$
Macroscopically, these amplitudes can be expressed
in terms of charge and quadrupole form factors of
the deuteron [10]:
$$
\Gamma_{00} =
               {\displaystyle \lim_{t \rightarrow 0} } ~
               [F_C(t) - { t \over {3M^2}} F_Q(t) ]
{}~~~,
\eqno{(15)}
$$
$$
\Gamma_{11} = \Gamma_{-1 -1} =
               {\displaystyle \lim_{t \rightarrow 0} } ~
               [F_C(t) + { t \over {6M^2}} F_Q(t) ]
{}~~~,
\eqno{(16)}
$$
where $F_C$ and $F_Q$ are measured in the units
of $e$ and $e/M^2$.
If the tensor combination of the amplitudes is taken, the first
terms cancel out and we obtain
the quadrupole term as
$ [\Gamma_{00}-{1 \over 2} (\Gamma_{11}+\Gamma_{-1-1}) ]/2
           = {\displaystyle \lim_{t \rightarrow 0} }
               -{t /{(4 M^2) }} F_Q(t) $.
Using this equation, we finally obtain the integral
as
$$
{\displaystyle \int} dx~ b_1^D(x)
         =
             {\displaystyle \lim_{t\rightarrow 0}}
              -{5 \over 3} {t \over {4M^2}} F_Q (t)
           +{1 \over 9} \delta Q_{sea}
{}~~~.
\eqno{(17)}
$$
This equation is very similar to the
Gottfried sum rule. If the sea is not $SU(2)_f$ symmetric,
the Gottfried sum rule is modified as
$$
{\displaystyle \int} dx~ [F_2^p(x)-F_2^n(x)]
         = {1 \over 3} + {2 \over 3} \int dx [\bar u(x)-\bar d(x)]
{}~~~.
\eqno{(18)}
$$
As we have the $SU(2)_f$ symmetric sea ($\bar U-\bar D=0$)
in a naive parton model,
the tensor polarization for sea quarks should vanish
($\delta Q _{sea}=0$) in the parton-model case.
Hence, we call the following equation
as a sum rule on the same
level with the Gottfried sum rule:
$$
{\displaystyle \int} dx~ b_1(x) =0~~~.
\eqno{(19)}
$$
If the sea quarks are tensor polarized, we obtain
a nonzero value
$$
{\displaystyle \int} dx~ b_1(x)
         = {1 \over 9} \delta Q_{sea}
{}~~~.
\eqno{(20)}
$$
All the results for $b_1(x)$ in Refs. 3, 12, and 13
satisfy the above sum rule in Eq. (19).
As the breaking of the Gottfried sum rule became
an interesting topic recently, it is
worth while investigating a possible mechanism to produce
the tensor polarization $\delta Q_{sea}$, which
breaks the sum rule.

Even though the sum-rule value is expected to be
zero for the $b_1$, it does not mean that $b_1$ itself
is zero. In fact, it is shown in the next section that
$b_1(x)$ can be negative in a certain $x$ region.

$~~~$

\noindent
{4. \underline{Examples of $b_1(x)$}

\skips

Some calculations for $b_1(x)$ are presented
in Refs. 3, 5, 11, 12, and 13.
We first discuss some examples
based on Ref. 3.
We consider the simplest case: a spin-1 system
with two spin-1/2 nucleons at rest.
This system obviously does not have a tensor
structure. Hence,
we have $b_1(x)=0$.

In the deuteron, a pion exchange produces a tensor
force between nucleons and gives rise to the
D-state admixture.
We use a convolution picture for calculating
the helicity amplitudes.
Namely, the helicity amplitude is given by
a helicity amplitude for the nucleon convoluted
with the light-like momentum distribution
of the nucleon.
$b_1$ for the deuteron is calculated as
$$
b_1(x)= \sum_{k=p,n} \int dy dz \delta (yz-x)
          [~ sin^2 \alpha ~ \Delta f_{dd} (y)
            -{{4 \sqrt 2} \over {\sqrt 5}}
             sin \alpha ~cos \alpha ~
                  \Delta f_{sd} (y)] F_1^k(z)
{}~~~,
\eqno{(21)}
$$
where $sin \alpha$ is the D-state admixture.
$\Delta f(y)$ is the light-cone momentum distribution
of a nucleon in the tensor polarized deuteron
[$\Delta f(y)=f^0(y)-(f^{+1}(y)+f^{-1}(y))/2$].
The first term $\Delta f_{dd}(y)$ is due to the D-state
and the second $\Delta f_{sd}(y)$ to the S-D interference.
Because of the small D-state admixture, the above
$b_1(x)$ is much smaller than the unpolarized $F_1(x)$.
(An extension of this work is done by
Khan and Hoodbhoy [11].)
The dynamics of producing the tensor structure
contributes to the nonzero $b_1(x)$.
However, it is
interesting to find that its integral still satisfies
the sum rule $\int dx b_1(x)=0$ in Eq. (19) by explicitly
integrating Eq. (21).

Miller studied an exchanged-pion contribution to
$b_1(x)$ [12]. Pions are associated with the tensor
force, so that it is natural to have large contributions
to $b_1$ from the pions.
The contribution is roughly parametrized as
$b_1(x)/F_1(x)\approx 0.02 (x-0.3)$.
If we integrate his pionic contribution (not the above
approximate equation), we find that the sum rule
$\int dx b_1(x)=0$ is still fulfilled.

Mankiewicz [13] studied $b_1(x)$ for the $\rho$ meson
by using light-cone wave-functions for constituent
quarks. Calculated $b_1(x)$ shape is very similar
to the one in Fig. 1.
$b_1(x)$ needs not be small in relativistic systems.

\vfill\eject
\vspace{-0.46cm}
\begin{list}{}{\leftmargin 0.0in \rightmargin 3.7in}
\item
$~~~~$
In order to illustrate how $b_1(x)$ looks like as
a function of $x$, we show the following example
[3].
We consider a relativistic system
with a quark with $j=3/2$ which
couples to another quark with $j=1/2$ to form a
$j=1$ state. In this case, $b_1$ is as large as
$F_1$ as shown in Fig. 1. The $b_1$ oscillates as
a function of $x$ and has negative values in the
medium-$x$ region. Integrating $b_1(x)$ over $x$,
we find that this example again satisfies
the sum rule $\int dx b_1(x)=0$.
\end{list}
\vspace{-1.0cm}
\hspace{6.6cm}
Fig. 1  $P_{3/2}$ quark coupled to a j=1/2 quark

\hspace{7.3cm}
(taken from Ref. 3).

We learned the following from the above examples.
Static nucleons alone do not contribute to $b_1$.
The dynamics of a pion exchange produces nonzero $b_1$.
$b_1(x)$ has very different $x$-dependence
from that of $F_1(x)$ or $g_1(x)$, and it satisfies
the sum rule $\int dx b_1(x)=0$ in all the cases
considered in this section.
$b_1$ is suitable for studying non-nucleonic
degrees of freedom in nuclei such as nuclear pions,
rhos, and perhaps nucleon substructures
if we find an experimental deviation from
conventional theoretical predictions.
Much physics could be studied in the near future,
for example, details of $b_1(x)$ for D, $^6$Li, $^{14}$N
and possible mechanisms of breaking the sum rule.

$~~~$

\noindent
{5. \underline{Conclusions}}

\skips

We discussed the tensor structure function $b_1$ based
on recent publications. Although much more theoretical efforts
have to be made to understand details of $b_1(x)$,
we expect that
$b_1$ provides important clues to physics of non-nucleonic
components in nuclei and to new tensor structures
on the quark-parton level.
Because $b_1$ for a nuclear target is considered to be small,
we need an intense electron accelerator for measuring it.
The proposed 15 GeV European Electron Facility is
an appropriate one for measuring $b_1$.

\vspace{0.7cm}

\begin{center}
{\underline{Acknowledgment}} \\
\end{center}

This research was supported by
the Deutsche Forschungsgemeinschaft (SFB 201).

$~~~$

\noindent
* E-mail: KUMANO@VKPMZP.KPH.UNI$-$MAINZ.DE.

\vfill\eject

\begin{center}
{\underline{References}} \\
\end{center}

\vspace{-0.30cm}
\vspace{-0.38cm}
\begin{description}{\leftmargin 0.0cm}
\item{[1]}
EMC Collaboration, J. Ashman et al.,
         Phys. Lett. {\bf B206}, 364 (1988);
for a summary,
R. L. Jaffe and A. Manohar,
         Nucl. Phys. {\bf B337}, 509 (1990).

\vspace{-0.38cm}
\item{[2]}
SMC Collaboration, research proposal at CERN (1988),
CERN/SPSC 88-47.

\vspace{-0.38cm}
\item{[3]}
P. Hoodbhoy, R. L. Jaffe, and A. Manohar,
         Nucl. Phys. {\bf B312}, 571 (1989);
see also R. L. Jaffe and A. Manohar,
         Nucl. Phys. {\bf B321}, 343 (1989);
P. Hoodbhoy, R. L. Jaffe, and E. Sather,
         Phys. Rev. {\bf D43}, 3071 (1991).

\vspace{-0.38cm}
\item{[4]}
There is a proposal to measure $b_1(x)$ at HERA:
HERMES Collaboration, research proposal at HERA (1989);
R. G. Milner,
in Electronuclear Physics with
                 Internal Targets,
                 edited by R. G. Arnold
                 (World Scientific, Singapore, 1989).

\vspace{-0.38cm}
\item{[5]}
A. Pais,
         Phys. Rev. Lett. {\bf 19}, 544 (1967);
L. L. Frankfurt and M. I. Strikman,
         Nucl. Phys. {\bf A405}, 557 (1983).

\vspace{-0.38cm}
\item{[6]}
F. E. Close and S. Kumano,
         Phys. Rev. {\bf D42}, 2377 (1990).

\vspace{-0.38cm}
\item{[7]}
B. L. Ioffe, V. A. Khoze, and L. N. Lipatov,
Hard Processes (North Holland, Amsterdam, 1984).

\vspace{-0.38cm}
\item{[8]}
NMC Collaboration,
P. Amaudruz et al.,
          Phys. Rev. Lett. {\bf 66}, 2712 (1991);
S. Kumano, Phys. Rev. {\bf D43}, 59 \& 3067 (1991);
S. Kumano and J. T. Londergan {\bf D44}, 717 (1991)
and references therein.

\vspace{-0.38cm}
\item{[9]}
R. P. Feynman, Photon Hadron Interactions
               (Benjamin, New York, 1972).

\vspace{-0.38cm}
\item{[10]}
M. Gourdin, Nuovo Cimento, {\bf 28}, 533 (1963);
F. E. Close,
             Phys. Lett. {\bf B65}, 55 (1976);
S. Kumano,
             Phys. Lett. {\bf B214}, 132 (1988);
             Nucl. Phys. {\bf A495}, 611 (1989).

\vspace{-0.38cm}
\item{[11]}
H. Khan and P. Hoodbhoy,
             Phys. Rev.  {\bf C44}, 1219 (1991);
             Phys. Lett. {\bf B298}, 181 (1993).

\vspace{-0.38cm}
\item{[12]}
G. A. Miller, in Electronuclear Physics with
                 Internal Targets,
                 edited by R. G. Arnold
                 (World Scientific, Singapore, 1989).

\vspace{-0.38cm}
\item{[13]}
L. Mankiewicz, Phys. Rev. {\bf D40}, 255 (1989).

\end{description}
\end{document}